Research paper

# Correlation Differential Power Analysis Attack to Midori64

*B. Khadem*[*1], *H. Ghanbari*[2], *A. Masoumi*[3]

1 Assistant Professor, Faculty of Computer Engineering, Imam Hossein Comprehensive University, Tehran.

2, 3 M.S.C. in Secure Communication and Cryptography, Imam Hossein Comprehensive University, Tehran.

**Extended Abstract**

**Background and Objectives:** Today, Internet communication security has become more complex as technology becomes faster and more efficient, especially for resource-limited devices such as embedded devices, wireless sensors, and radio frequency identification (RFID) tags, and Internet of Things (IoT). Lightweight encryption algorithms provide security for these devices to protect data against intruders. A complete understanding Lightweight encryption helps people find better ways to protect valuable information as technology advances faster. But the limitation of using energy in lightweight block ciphers (LBCs) is one of the major challenges for ever-expanding IoT technologies. Also, these LBC are subject to numerous attacks. Side-channel attacks are among the most cited threats to these ciphers. In these attacks, the attacker benefits from a physical leak of the cryptographic chip to reach the cryptographic key. A type of side-channel attack is power analysis in which the attacker reaches the key using the relationship between the power consumption of the chip during the algorithm running, data processing, and operations.

**Methods:** A popular LBC that has been recently introduced in the IoT is Midori. In this paper, a differential power attack (DPA) to the Midori64 block cipher is designed. According to the proposed method, an attack on the S-boxes of the first round is done to obtain half of the master key bits. Then, the S-boxes of the second round were attacked to obtain remaining the master key bits.

**Results:** The results confirmed that the key is ultimately obtained in this cipher by performing an attack, so the block cipher is not resistant to a DPA. With the low volume of computational complexity, we obtained the Midori block cipher key, which was considered secure, just by using 300 samples of the plaintext.

**Conclusion:** Following the running of Midori64 on the AVR microcontroller of the Atmega32 model, the master key of Midori block cipher is discovered with 300 known texts. Furthermore, we obtained the master key with a smaller number of samples than the electromagnetic analysis attack.

**Introduction**

Recently, numerous security threats have emerged along with advancements in IoT technologies [1], which makes encryption as an important step to protect systems against these threats. So, low power ciphers that can be used in IoT devices have attracted the attention of many researchers. However, it is critical to use LBC, as we are encountering budget limitations in consuming power and utilizing resources. Numerous studies have so far been conducted on energy-efficient LBCs [2-6]. Among the available LBCs, the Midori exhibits the least power consumption [7]. This feature multiplies the importance of evaluating the security of this block cipher. In the area of hardware security, any cryptography block cipher, including Midori, is subjected to many potential side-channel threats [8]. In this attack, secret cryptographic

Doi:

data (e.g., master key) are accessible by using the side leak analysis of the chip while performing encryption. These leaks include power consumption leaks, electromagnetic leaks, and algorithm running time leaks. This paper aims to investigate the security of the Midori block cipher against DPA, as well as information leakage from this block cipher during encryption operations.

A. Literature review

Following the emergence of the Midori block cipher and the investigation of some associated theoretical attacks [2], many studies have been conducted to analyses the security of this block cipher against potential attacks. The low power consumption of the Midori block cipher seems to be the main reason for researchers to work with this block cipher, as saving resources and power consumption is a critical factor to consider today.

The first study on the power analysis attack to the Midori block cipher was conducted by Heuser et al. in 2017 [9]. They examined the resilience of a few other lightweight ciphers against power analysis attacks. For this, non-profiled and profiled side-channel attacks were arranged on the first, the last, or both rounds simultaneously. In the non-profiled attack on the first round of ciphers AES, ZORRO, Robin, LED, Midori64, Mysterion, KLEIN, Piccolo, PRESENT, PRIDE, RECTANGLE, and Skinny, no difference was observed between their 8x8 substitution boxes (S-boxes) in resisting against the power analysis attack. However, the 4-bit Midori64 and KLEIN S-boxes showed more resilience to power analysis attacks than the 8-bit boxes. The same attacks were performed using the profiled method and machine learning, and it is found that also, attacking 4-bit S-boxes is easier than 8-bit S-boxes.

There are many other researches in DPA, which some of them are described here.

In FSE 2005, transparency order was proposed by Prouf [10] as a parameter for the robustness of S-boxes to DPA. Prouf rewrote DPAs in terms of correlation coefficients between two Boolean functions. He exhibited properties of S-boxes (also called (n, m)-functions) relied on DPA attacks. He showed that these properties were opposite to the non-linearity criterion and to the propagation criterion. To quantify the resistance of an S-box to DPAs, he introduced the notion of transparency order of an S-box and studied this criterion with respect to the non-linearity and to the propagation criterion.

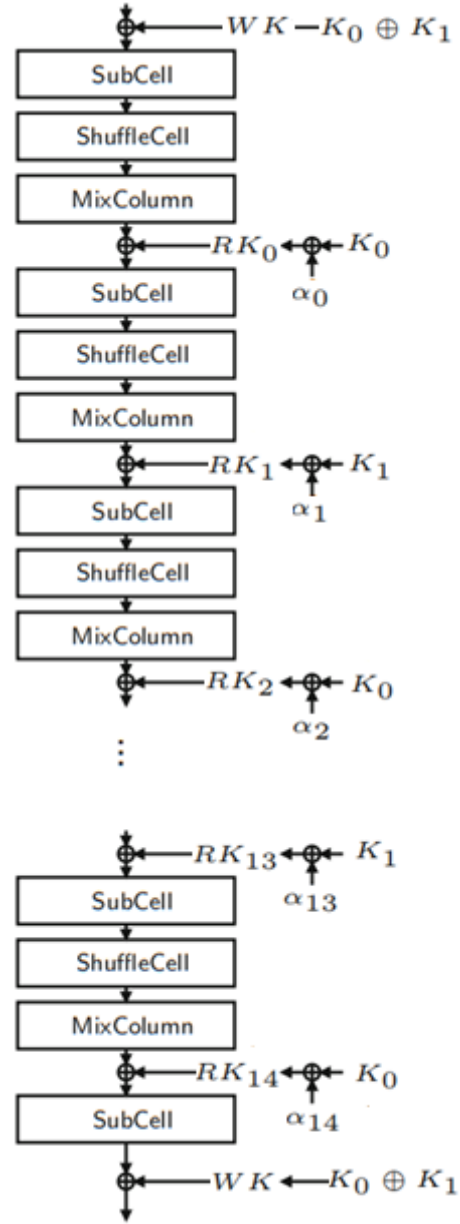

Fig. 1: Midori64 block cipher block diagram [2]

Carlet (2005) [11] found that If transparency order has sufficiently small value, then the S-box is able to withstand DPA attacks without that ad-hoc modifications in the implementation be necessary (since these modifications make the encryption about twice slower). Also he proved lower bounds on the transparency order of highly nonlinear S-boxes. He showed that some highly nonlinear functions (in odd or even numbers of variables) have very bad transparency orders.

Fei et al. (2012) [12] proposed a statistical model for DPA that takes characteristics of both the physical implementation and cryptographic algorithm into consideration. Their model established a quantitative relation between the success rate of DPA and a cryptographic system. The side channel characteristic of the physical implementation was modeled as the ratio between the difference-of-means power and the standard deviation of power distribution. The side channel property of the cryptographic algorithm was extracted by a novel algorithmic confusion analysis. Experimental results on DES and AES verified this model and demonstrate the effectiveness of algorithmic confusion analysis.

Picek et al. (2014) [13] employed a novel heuristic technique to generate S-boxes with "better" values of the confusion coefficient in terms of improving their side-channel resistance. They conducted extensive side channel analysis and detect S-boxes that exhibit previously unseen behavior. For the 4 × 4 size they found S-boxes that belong to optimal classes, but they exhibited linear behavior when running a CPA attack, therefore preventing an attacker from achieving 100% success rate on recovering the key.

Parthiba et al. (2018) [14] presented the use of Differential Cascade Pre-resolve Adiabatic logic (DCPAL) based adiabatic circuits for the implementation of cryptosystems to prevent the DPAs on chips. The DCPAL is an adiabatic logic, which incurs lower power consumption and earns its application in the design of low power cryptosystems. The property of power analysis resistance was demonstrated through the use of DCPAL implementation for an S-box. The S-box was implemented in both the standard CMOS and the DCPAL styles to prove the power analysis resistance and low-power operation capability. Fair comparisons had been made for validation. The advantage of using the DCPAL for DPA resistant systems was also demonstrated through the S-box implementation. Extensive transient simulations had been carried out using the technology files from 180 nm foundry.

Carlet et al. (2020) [15] by using an optimal distinguisher under an additive Gaussian noise assumption, clarified how an attacker can make side-channel attacks as difficult (resp., easy) as possible, in relation with the auto-correlation spectrum of Boolean functions. They then constructed balanced Boolean functions that are optimal for each of these two scenarios. Generalizing the objectives for an S-box, they analyzed the auto-correlation spectra of some well-known S-box constructions in dimensions at most 8 and compared their intrinsic resiliency against side-channel attacks. Finally, they performed several simulations of side-channel attacks against the aforementioned constructions, which confirm their theoretical approach.

In hardware security, side-channel attacks pose a danger because they illegally analyze the secret key in a cryptographic device using the power consumption and electromagnetic waves generated during the device's operation. One type of side-channel attack that uses electromagnetic waves is called electromagnetic analysis. For the first time, the Midori block cipher was attacked by electromagnetic analysis in 2017, where the key got accessible using 18,000 samples by Yoshikawa et al. [7] who proposed a method of electromagnetic analysis.

In addition to hardware attacks, other attacks have been reported on the Midori64 block cipher, some of which are described here.

Gérault et al. (2016) [16] proposed a constraint programming model to automate the search for optimal (in terms of probability) related-key differential characteristics on Midori. Using it, they built related-key distinguishers on the full-round Midori64 and Midori128, and mount key recovery attacks on both versions of the cipher with practical time complexity, respectively $2^{35.8}$ and $2^{43.7}$.

Guo et al. (2016) [17] presented an invariant subspace attack on the block cipher Midori64. Their analysis showed that Midori64 has a class of 232 weak keys. Under any such key, the cipher can be distinguished with only a single chosen query, and the key can be recovered in 216 time with two chosen queries. As both the distinguisher and the key recovery have very low complexities, they confirmed their analysis by implementing the attacks. Some tweaks of round constants made Midori64 more resistant to the attacks, but some guided to even larger weak-key classes. To eliminate the dependency on the round constants, they investigated alternative S-boxes for Midori64 that provide certain level of security against the found invariant subspace attacks, regardless of the choice of the round constants.

Chen et al. (2017) [18] proposed impossible differential cryptanalysis of Midori64. They studied the non-linear layer of the cipher and give two useful properties. They also found the first 6-round impossible differential paths with two non-zero and equal input cells and one non-zero output cell, and then mount 10-round attack. This was the first impossible differential attack on Midori.

Shahmirzadi et al. (2018) [19] studied security of Midori64 against impossible differential attack. To this end, they used various techniques such as early-abort, memory reallocation, miss-in-the-middle and turning to account the inadequate key schedule algorithm of Midori64. They first showed two new 7-round impossible differential characteristics which were the longest impossible differential characteristics found for Midori64. Based on the new characteristics, they mounted three impossible differential attacks on 10, 11, and 12 rounds on Midori64 with $2^{87.7}$, $2^{90.63}$, and $2^{90.51}$ time complexity, respectively, to retrieve the master-key.

Todo et al. (2019) [20] introduced a new type of attack, called nonlinear invariant attack, and tested it on block ciphers Scream, iScream, and Midori64 in a weak-key setting to confirm its accuracy. They reported that if the Midori64 master key is set weakly, the master key is reached by holding a large number of pairs of the original plaintext and ciphertext.

Tim Beyne (2020) [21] developed a new approach to invariant subspaces and nonlinear variations. His results showed that with minor modifications to some of the round constants, Midori-64 shows a nonlinear invariant with 2^96+2^64 corresponding weak keys. By combining the new invariant with integral cryptanalysis, a practical key-recovery attack on ten rounds of unmodified Midori-64 is obtained. The attack works for 2^96 weak keys and irrespective of the choice of round constants.

Ru et al. (2020) [22] carried out the key invariant deviation and linear analysis of Midori-64 using the method of related-keys technology and linear analysis. According to the nature of the key extension algorithm of Midori-64, a suitable input-output linear mask and key difference were selected to construct a 7-round related-keys invariant linear discriminator. The data complexity and time complexity of the attack were respectively 2^62.99 and 2^76.6.

There are further studies on the threshold implementation of the Midori block cipher to strengthen it against potential attacks [23].

However, to the best of our knowledge, no studies have reported on DPA against Midori64 block cipher which is proposed in this paper and implemented on the AVR microcontroller.

*B. Midori: An energy-efficient block cipher*

Midori is known as LBC with low power consumption [2]. The structure of this block cipher is provided in two 128-bit and 64-bit versions. In both versions, the length of the key is 128 bits (Figure 1). Midori64 and Midori128 respectively consist of 16 and 20 rounds respectively.

In this paper, we considered Midori64. Initially, 64-bit blocks are placed inside a 4×4 matrices called the state matrix (S matrix). Each entry of this matrix is 4 bits in length. All subsequent processing are applied to the entries of this matrix and updated it.

In Midori64, the XOR operation is done on a 128-bit key with a 64-bit block. The length of the 128-bit key is divided into two 64 bit keys $k_0$ and $k_1$ (K= $k_0||k_1$). In the first round, WK is formed according to Equation (1). The sub-keys of the next rounds ($Rk_1$), which are obtained from the secret key K, were obtained from Equation (2).

$$WK = k_0 \oplus k_1 \quad (1)$$

$$RK_{r-1} = K_{(r-1 \bmod 2)} \oplus \alpha_{r-1} \quad (2)$$

In Equation (2), $\alpha$ is an invariant value per each round.

As shown in Figure 1, in the first round, the value of WK is XORed with the values of the S matrix. After updating, the values of the S matrix enter the substitution box (SubCell). The S-box for both the 64-bit and 128-bit versions of Midori block cipher is a 4-bit and is showed in Table 1.

Table 1. Midori64 block cipher S-box (y=S(x))

| x | 0 | 1 | 2 | 3 | 4 | 5 | 6 | 7 | 8 | 9 | A | B | C | D | E | F |
|---|---|---|---|---|---|---|---|---|---|---|---|---|---|---|---|---|
| y | C | A | D | 3 | E | B | F | 7 | 8 | 9 | 1 | 5 | 0 | 2 | 4 | 6 |

After passing the S-box and updating the S matrix, these values enter the Shuffle-Cell stage and are passed through the permutation (Table 2).

Table 2. Shuffle-Cell (W=Sh(i))

| i | 0 | 1 | 2 | 3 | 4 | 5 | 6 | 7 | 8 | 9 | A | B | C | D | E | F |
|---|---|---|---|---|---|---|---|---|---|---|---|---|---|---|---|---|
| w | 0 | 7 | E | 9 | 5 | 2 | B | C | F | 8 | 1 | 6 | A | D | 4 | 3 |

The S matrix values enter the Mix-Column stage and are updated using the M matrix and Equation (3) the S matrix is updated and enter the next stage. These stages last up to 15 rounds. The output of the 15th round after XOR with the corresponding round key enters the S-box for the next round. Then, the output of this S-box is XORed repeatedly with the sub-key of the first round (WK).

$$(s_i, s_{i+1}, s_{i+2}, s_{i+3})^T \leftarrow M(s_i, s_{i+1}, s_{i+2}, s_{i+3})^T \quad (3)$$

$$i \in (0,4,8,12)$$

$$M = \begin{pmatrix} 0 & 1 & 1 & 1 \\ 1 & 0 & 1 & 1 \\ 1 & 1 & 0 & 1 \\ 1 & 1 & 1 & 0 \end{pmatrix}$$

*C. Power analysis attack*

Power analysis is a way to obtain secret data using the power consumption of a chip during running an encryption program. Power analysis is based on the fact that the instantaneous power consumption of the chip depends on the data under processing and the performing operations. This attack includes two types of simple power analysis (SPA) attack and DPA [8]. In DPA, using equation 4, a correlation is formed between the power matrix measured from the chip during the encryption operation and the hypothetical power matrix, which uses the Hamming weight of the data after XOR with the key and passage through the encryption box. The result will be into a matrix r. The column containing the largest value of this matrix is equal to one byte of the key.

$$r_{i,j} = \frac{\sum_d^D (h_{d,i} - \bar{h}_i) \cdot (t_{d,j} - \bar{t}_j)}{\sqrt{\sum_d^D (h_{d,i} - \bar{h}_i)^2 \cdot \sum_{d=1}^D (t_{d,j} - \bar{t}_j)^2}} \quad (4)$$

Where h is a hypothetical power matrix that is obtained by using the Hamming weight of data for different hypothetical keys, t is the matrix of measured powers, $\overline{t_j}$ and $\overline{h_j}$ are the mean values of the matrices t and h, respectively, and D is the number of power samples obtained for different known plaintexts (D known different plaintexts).

Figure 2 shows the steps of running a DPA. The central issue in a DPA attack is to choose the right location to measure the power and run the attack. In block ciphers, according to the implementation type and structure of the S-box, the power measurement location is usually after the S-box of the first round.

The remainder sections of this paper are organized as follows. In section 2, a brief description of the structure of the Midori block cipher and the power analysis attack is presented. Section 3 proposes a method for performing power analysis attack on Midori. In section 4, the experimental results of the attack are declared and the accuracy of the method presented in section 3 is confirmed. The results are ultimately presented in section 5.

*D. Innovation*

In this paper, a DPA is performed for the first time on the version of the Midori64 block cipher implemented on the AVR microcontroller. Another innovation of this research is improved efficiency of DPA to the Midori64 block cipher to reach the master key using less sample number and lower computational complexity related to [9].

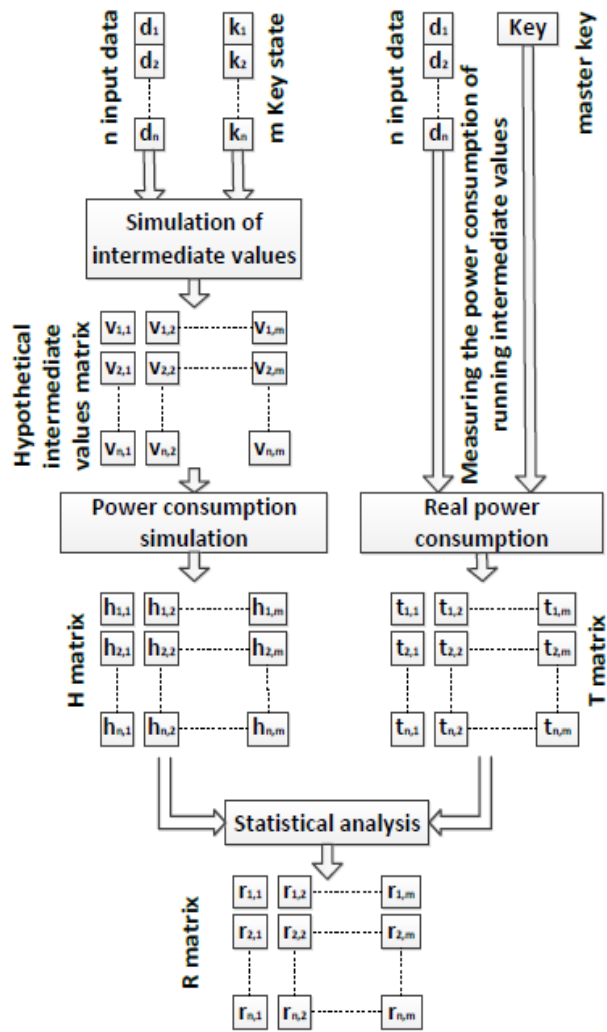

Fig. 2. Steps to DPA implementation and the formation of matrix r

**Attack design**

As described in Section 1.2, the input block of the Midori block cipher enters a 4x4 matrix and all modifications are applied to it. Each entry of this matrix is 4 bits in length that pass through 4-bit S-boxes. Therefore, there are 16 S-boxes in this block cipher. To obtain the key, operations are done within a 4-bit to 4-bit plan, and in each step, we obtain 4 bits of the key (Figure 3). As shown in Figure 3, the output Hamming weight of the S-box is used as a power simulation model to obtain the key.

Since the S-boxes of this block cipher are 4-bit in length, the key is obtained within a nibble-to-nibble design. Each nibble has 16 different modes, for all of which the output of the S-box is guessed and its Hamming weight is determined.

If the guess is correct, one of the values of one column of the matrix that form the Pearson correlation equation [4] will be greater than the others. As a result, the column number that indicates this value is equal to 4 bits of the favorite key (Figure 4). Figure 4 shows the point of the attack on a S-box of a block cipher.

Four bits of the key are obtained following the above operations. For the remaining 60 bits of the key, the above steps should be repeated 15 more extra times. Therefore, a total of $2^4 \times 16 = 256$ calculations must be accomplished.

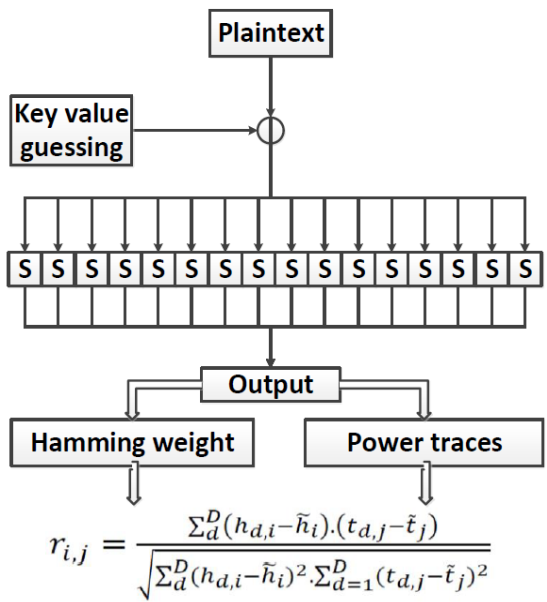

Fig. 3. The point of attack Assuming the key value (16 various mods)

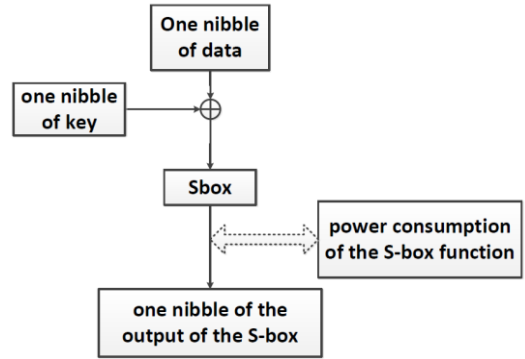

Fig. 4. The point of attack on one of the S-boxes

Importantly, these 4 bits are not the whole master key but show the WK value. After obtaining the value of WK, in the second stage, an attack must be performed on the second round of the block cipher to obtain the value of $k_0 \oplus \alpha_0$. After obtaining this value and since the value $\alpha_0$ is known, the value $k_0$ and consequently the value of the master key will be obtained. Since both the plaintexts and cipher texts are identified in each block cipher, the best point to perform the attack is the first or last round. The Midori block cipher has also been attacked electromagnetically[7].

Test results

The proposed method is practically implemented in this section.

  A. Test layout

The tools employed in experiments are introduced in Figure 5 and Table 3. To perform the attack, we run the Midori64 block cipher on the board with the AVR chip (model atmega32). A 1-Ω resistor is installed on the board to determine the power consumption of the chip during encryption at the base of the block cipher output from the microcontroller. After passing through this resistor, the output current from the microcontroller is stored in the oscilloscope according to equation P=RI2 in terms of power consumption. The SMA connector is used on the board to reduce the noise.

After writing the C code of the Midori block cipher on the chip, the board is connected to the oscilloscope and the computer, and plaintext is randomly sent to the board via the computer. After encryption by the chip, the encrypted values are sent to the computer. If there is conformity between cipher text from the board and the computer, the computer sends a command to the oscilloscope to store the measured power values.

Following these operations, the values stored on the oscilloscope, which are into CSV files, are transferred to the computer by flash. Then, the DPA attack is done on data in MATLAB 2016b according to equation (4) and the steps described in section 1.3.

Table 3. Experimental tools

| block cipher algorithm | Midori |
|---|---|
| Block length | 64 bits |
| Board used | Predesigned board with a micro AVR |
| Microcontroller | AVR atmega32 |
| Development tool | Code vision AVR |
| Oscilloscope | Key sight Infiniium (model DSO9064A) |
| Sampling rate | 1Gsa/sec |
| Computer | HP pavilion |
| Memory | 8GB |
| CPU | Intel core i7 5500 |
| Software | Matlab 2016b |

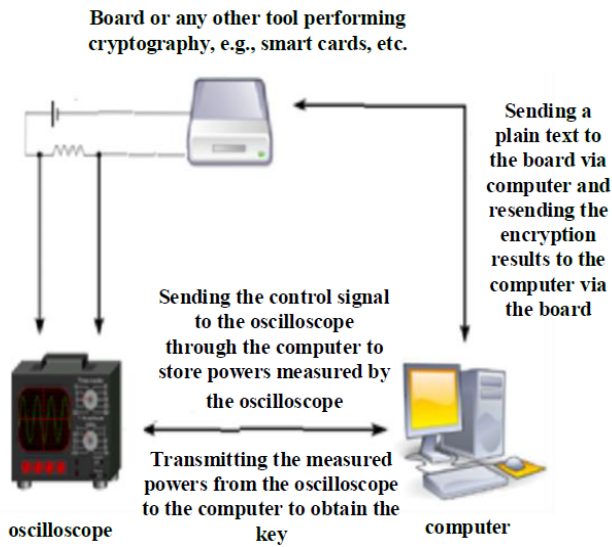

Fig. 5. Laboratory tools to perform a DPA attack on the Midori block cipher

*B. Experimental results*

A challenge when attacking 4-bit S-boxes is that there are only 16 input modes that makes it more difficult to find the key. To fix this problem, we made entries to the cipher more randomly to more modify the registers inside the microcontroller and increase power leakage. The input data is 4-bit numbers (from zero to 15) generated in MATLAB. Since the rand function in MATLAB generates random numbers as a uniform distribution, the input distribution is uniform.

At the end of the experiment, we obtained 4 bits of WK with 300 input samples (Figure 6). In this experiment, we set 4 bits of WK to 10, and according to Figure 6, the graph is higher than the other numbers on the number 10, which shows 4 bits of WK.

However, the oscilloscope is capable of pre-processing power samples. For each known input plaintext, the oscilloscope repeated the experiment 256 times and stored the data after averaging the obtained power samples. By which, the noise is greatly A secondary section heading is enumerated by a capital letter followed by a period and is flush left above the section. The first letter of each important word is capitalized and the heading is italicized.

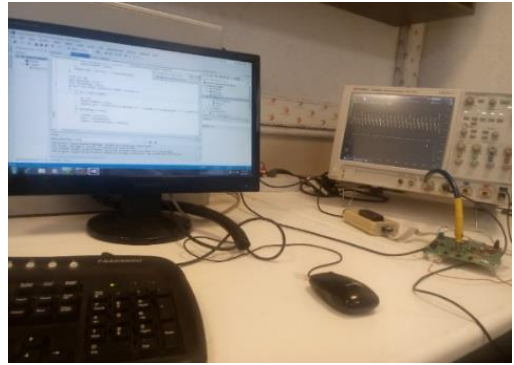

Fig 6. The board-oscilloscope-computer connection to perform DPA attack Board or any other tool performing cryptography, e.g., smart cards.

*C. Analysis of experimental results*

Figure 7 shows the correlation diagram of the hypothetical and real power consumption. As discussed earlier, the Hamming weight model is used in this attack to model hypothetical power consumption. The horizontal axis of this diagram shows the numbers 1 to 16, which are all types of keys, where we calculate the Hamming weight of the S-box for each of them according to a given input.

*B. Khadem, H. Ghanbari, A. Masoumi*

A matrix is then formed which enters the Pearson equation (Equation (4)) with the power consumption matrix for the same input samples. Figure 6 shows a correlation diagram between power consumption and hypothetical power modeled by the Hamming weight. In this diagram, the greatest correlation between the two matrices occurs in exchange for the correct key guess, which causes the maximum diagram at point 10. In simple terms, most of the correlation between the hypothetical power calculated with the key 10 by the Hamming weight model and the power consumption of the cryptographic chip occurred in the value of 10.

### Results

This paper is carried out to perform a DPA attack on the Midori64 block cipher. According to the proposed method, an attack on the S-boxes of the first round is done to obtain WK. Then, the S-boxes of the second round were attacked to obtain $k_0$ and the master key. The key is ultimately obtained in this cipher by performing an attack. The results confirmed that the block cipher is not resistant to a DPA attack. With the low volume of computational complexity, we obtained the Midori block cipher key, which was considered secure, just by using 300 samples of the plaintext. Furthermore, we obtained the master key with a smaller number of samples than the electromagnetic analysis attack [7]. It is recommended to use the mask method and its effect on the DPA attack, as well as the deep learning method in future studies.

### Acknowledgment

The authors want to thank Mr. Mohammad Gholi, Mr. Habibi, and Mr. Mahdavi for their kind contributions during this paper.

### Biographies

**BEHROOZ KHADEM** received the B.Sc. degree in applied mathematics from the University of Tehran, Tehran, Iran, in 1991, the M.Sc. degree in applied mathematics from the Shaheed Bahonar University of Kerman, Kerman, Iran, in 1995, and the Ph.D. degree in chaos-based cryptography from the Department of Mathematics, Kharazmi University, Tehran, in 2015. Since 2011, he has been a Research Assistant at the Institute of Mathematics, Kharazmi University. He has published over 50 research articles in reputed peer-reviewed journals and conference proceedings of IEEE/Springer/Elsevier. His research interests include data security and cryptography, but are not limited to applied mathematics, algorithms and complexity, computer security, chaos-based cryptography, applied cryptanalysis, security of communication networks, arti_cial intelligence and machine learning for **HAMID GHANBARI** received the B.Sc. degree in IRIB University, Tehran, Iran, in 2019, and the M.Sc. degree in electrical engineering-telecommunication (cryptography and secure communication) from Imam Hossein University (IHU), Tehran, Iran, in 2020. His research interests include hardware implementation of cryptographic algorithms and side channel analysis.

**AMIN MASOUMI SUTEH** received the B.Sc. degree in power engineering from the Mazandaran University of Science and Technology (MUST), Babol, Iran, in 2015, and the M.Sc. degree in electrical engineering-telecommunication (cryptography and secure communication) from Imam Hossein University (IHU), Tehran, Iran, in 2018. He was the Secretary of the Iranian Academic Society of Cryptology, Students' Branch, IHU, in 2017. His research interests include security protocols and provable security models.

security, image processing, and optimization techniques. He has served as a reviewer and a technical program committee member of multiple journals and conferences.

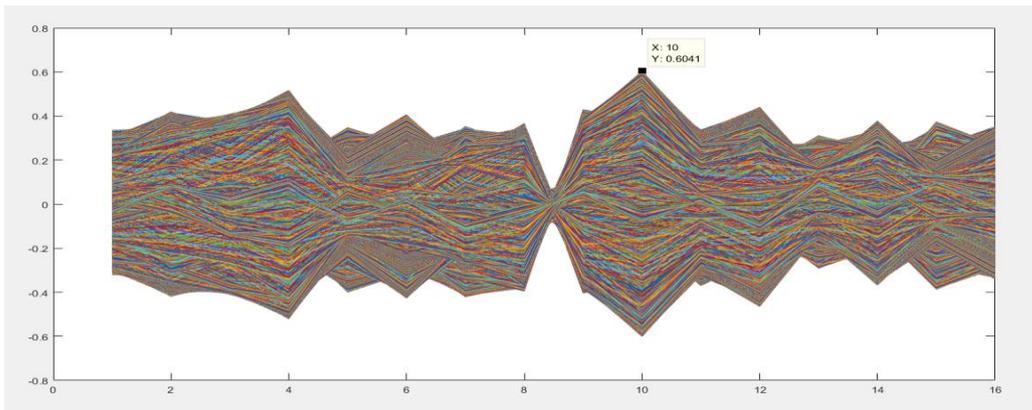

Fig. 7. The results of implementing a DPA attack on the power data obtained from an oscilloscope in Matlab (Key 10 is obtained)